# Fast and Accurate Modeling of Transient-state Gradient-Spoiled Sequences by Recurrent Neural Networks

Hongyan Liu[1], Oscar van der Heide[1], Cornelis A. T. van den Berg[1], Alessandro Sbrizzi[1]

**Corresponding Author:** Hongyan Liu, University Medical Center, Heidelberglaan 100, 3508 GA Utrecht, The Netherlands. E-mail: h.liu@umcutrecht.nl

[1] Center for Image Sciences, University Medical Center Utrecht, Utrecht, The Netherlands

# Abstract

Fast and accurate modeling of MR signal responses are typically required for various quantitative MRI applications, such as MR Fingerprinting and MR-STAT. This work uses a new EPG-Bloch model for accurate simulation of transient-state gradient-spoiled MR sequences, and proposes a Recurrent Neural Network (RNN) as a fast surrogate of the EPG-Bloch model for computing large-scale MR signals and derivatives. The computational efficiency of the RNN model is demonstrated by comparing with other existing models, showing one to three orders of acceleration comparing to the latest GPU-accelerated open-source EPG package. By using numerical and in-vivo brain data, two use cases, namely MRF dictionary generation and optimal experimental design, are also provided. Results show that the RNN surrogate model can be efficiently used for computing large-scale dictionaries of transient-states signals and derivatives within tens of seconds, resulting in several orders of magnitude acceleration with respect to state-of-the-art implementations. The practical application of transient-states quantitative techniques can therefore be substantially facilitated.

# Abbreviations

| | |
|---|---|
| **AD** | Automatic differentiation |
| **CRLB** | Cramér-Rao lower bound |
| **DE** | Differential Evolution |
| **EPG** | Extended Phase Graph |
| **GPU** | Graphics Processing Unit |
| **GRU** | Gated Recurrent Unit |
| **MAE** | Mean Absolute Error |
| **MAPE** | Mean Absolute Percentage Error |
| **MR-STAT** | Magnetic Resonance Spin Tomography in Time-Domain |
| **MRF** | Magnetic Resonance Fingerprinting |
| **NRMSE** | Normalized Root Mean Square Errors |
| **RNN** | Recurrent Neural Network |
| **qMRI** | Quantitative Magnetic Resonance Imaging |
| **SSFP** | Steady-State Free Precession |

# 1 Introduction

Quantitative Magnetic Resonance Imaging (qMRI) aims at reconstructing various magnetic properties of tissues, such as relaxation times and proton density. Such quantitative tissue parameter maps may reveal pathological information about organs that are theoretically independent of MR protocols, and therefore they may lead to more objective and precise clinical diagnosis.[1,2] Recent qMRI methods, such as MR Fingerprinting (MRF)[3] and MR-STAT[4], usually use relatively short sequences, during which the magnetization is in a transient state, to encode multiple quantitative parameters into the measured signal simultaneously. To perform the quantitative reconstructions, the spin dynamics needs to be simulated according to the physical model of MR signal, for example, the Bloch Equation or the Extended Phase Graph (EPG).[5]

Currently gradient-spoiled transient-state MR sequences are widely used for qMRI applications[6–8], because they require relatively short acquisition times, and are not affected by the banding artifacts resulting from the main ($B_0$) field non-uniformity. For these gradient-spoiled sequences, the EPG formalism is an efficient computational model to simulate the MR signal evolution. Unlike the Bloch equation which models the temporal MR signal for a single isochromat resonating at a given frequency, the EPG approach models the spins within a voxel as a discrete set of dephasing states, and is an efficient way for simulating spin dephasing induced by unbalanced spoiling gradients.

For various qMRI applications, computing large amounts of MR signals using the EPG models are necessary but also very time-consuming tasks. For example, in MRF methods, a dictionary which contains the temporal signal evolution for different combinations of parameter values ($T_1, T_2, B_1^+$, etc.) is usually used for reconstructing pixel-wise properties by a Matched Filter estimator (dictionary matching)[9]. To reconstruct all parameters with high precision, many discretization steps are required for each parameter to be estimated, resulting into a large dictionary which may need prohibitively long time, e.g., hours or even days, to generate[10]. Furthermore, modification of the sequence acquisition parameters asks for new dictionary computations.

Recently, progress in accelerating MR signal simulations has been made mainly in two different directions. On one hand, algorithms for large-scale MR signal simulations, based on the exact MR physics models, have been implemented and executed in parallel on graphics processing units (GPU), and achieve reasonable acceleration rate[11,12]. On a parallel track, based on the recent development in the deep learning fields, several types of neural networks (NN), such as the generative adversarial networks (GAN)[13] and the artificial neural networks (ANN)[14], have been trained as surrogate models to compute MR signals at very high speed. However, one common problem for these neural network methods is that they all are currently trained and validated for a fixed MR sequence and allow for limited, if any, sequence parameter inputs such as TE and TR. Therefore, these models have very little flexibility and have to be re-trained when a new or slightly modified sequence is applied.

Besides simulating the temporal MR signals, qMRI applications sometimes also require computing the MR signal derivatives with respect to the reconstruction parameters. For example, derivative computations are usually required for gradient-based optimization algorithms in order to solve model-based qMRI problems[4,15]. One other example requiring derivative computations is the Cramér-Rao-based optimal experimental design framework for qMRI sequences[16,17]. Signal derivatives can be computed using either finite difference (FD) or automatic differentiation (AD) method[16–18], which may at least double the amount of calculation compared to computing only the signal without derivatives.

In this work, we focus on how to accurately and efficiently model MR signal responses for gradient-spoiled sequences. Firstly, a more accurate EPG model, namely EPG-Bloch, is proposed. Currently, widely used EPG models either consider the radiofrequency (RF) pulses as instantaneous spin rotations[6] or correct for the RF slice profile imperfections using an approach that is based on the small tip-angle approximation[19–21]. To better model the imperfect slice profile effects, especially for large flip-angles (>60 degrees), the new EPG-Bloch model considers the RF pulse shape and models the RF excitation effects by temporal discretization and sub-step evolution, analogously to Bloch-based simulation of RF excitation pulses[22,23]. Experimental results show that signals computed by EPG-Bloch model fit better to measured signals comparing to conventional EPG model signals. Our findings are thereby in agreement with another recently published work addressing a similar problem[24]. In Reference[24], the slice profile effects are modelled by a frequency-domain analysis of the RF effects in the EPG framework while here we propose a spatial domain approach.

The here proposed EPG-Bloch model is computationally more expensive than the standard EPG because it requires simulating both the signal for many dephasing states and the discretization of RF pulses. The long computational time of the EPG-Bloch motivates the usage of a fast surrogate model. In order to significantly reduce the runtime, we apply one of the most widely used deep learning models for time series prediction, the recurrent neural networks (RNN)[25–27], as a surrogate model for EPG-Bloch. The RNN model takes both tissue parameters and sequence parameters as inputs, and outputs the MR signals as well as their derivatives sequentially. The sequence parameters could be time-varying such as a transient-state flip-angle scheme[6–8]. Several deep learning methods have been applied for accelerating the computation of large-scale MR signal evolution. Comparing to existing network architectures[13,14], the proposed RNN network has the following advantages. Firstly, once trained, the same RNN can be used for sequences with various sequence parameters, such as sequence length $N_{TR}$ and time-dependent flip-angle train $\boldsymbol{\alpha}$; therefore, training of the RNN model can be a one-time task for various experimental setups. Secondly, the RNN model is a structured deep neural network which recurrently uses the same units, thus the number of parameters (~20000) in the network and the size of the training set are relatively small. Finally, many deep learning techniques have been developed to optimize RNN computation on both CPUs[28] and GPUs[29,30], and this makes fast and parallelizable computation of the RNN model possible with existing publicly available software implementations.

We demonstrate that the RNN model could be trained and used as a fast and accurate surrogate model for EPG-Bloch with an appropriate training dataset, and it can be conveniently used for large-scale MR signal simulations for different acquisition parameters without need for re-training. Overall, we show that the RNN model on GPU accelerates the signal computation by at least a factor of $10^4$ compared to the EPG-Bloch model, and is approximately one to three orders of magnitude faster than the state-of-art GPU- accelerated EPG simulation package[11] for different sizes of dataset. By using numerical and in-vivo experimental data we provide examples of applications of RNN for rapid MR signal computations, namely (1) the generation of MRF dictionaries and (2) the optimal sequence design based on Cramér-Rao lower bound (CRLB). In both scenarios, construction of a 3-dimensional dictionary and optimization of a 3 second long sequence, the total computation times are within a few seconds.

# 2 Theory

## 2.1 EPG-Bloch

The extended phase graph (EPG) represents the MR signal obtained from a voxel of volume $V$ under the effect of dephasing terms using the Fourier formalism as follows[5]:

$$\tilde{F}_+(\mathbf{k}) = \int_V \left( M_x(\mathbf{r}) + iM_y(\mathbf{r}) \right) e^{-i\mathbf{k}\mathbf{r}} d\mathbf{r} = \int_V M_+(\mathbf{r}) \, e^{-i\mathbf{k}\mathbf{r}} d\mathbf{r},$$

$$\tilde{F}_-(\mathbf{k}) = \int_V \left( M_x(\mathbf{r}) - iM_y(\mathbf{r}) \right) e^{-i\mathbf{k}\mathbf{r}} d\mathbf{r} = \int_V M_-(\mathbf{r}) \, e^{-i\mathbf{k}\mathbf{r}} d\mathbf{r},$$

$$\tilde{Z}(\mathbf{k}) = \int_V M_z(\mathbf{r}) \, e^{-i\mathbf{k}\mathbf{r}} d\mathbf{r},$$

$$(1)$$

where $\mathbf{k}$ is the dephasing coordinate for the "configuration state" $\tilde{\mathbf{F}}(\mathbf{k}) = [\tilde{F}_+(\mathbf{k}), \tilde{F}_-(\mathbf{k}), \tilde{Z}(\mathbf{k})]^T$. For gradient-spoiled sequences, the dephasing term mainly comes from spoiler gradients.

In Reference[5], it is shown that the evolution of configuration states for a given MR sequence can be computed by applying different physical operators, such as RF excitation, relaxation and dephasing. To allow for more accurate computations, slice profile correction can be included by discretizing the slice-selective dimension $z$ into $N_z$ sub-slices, and computing the evolution of these spatial-dependent configuration states, $\tilde{\mathbf{F}}(\mathbf{k}, z)$, separately[19,21]. The slice profile can be computed by small tip-angle approximation[31], and then multiplied with the original flip-angle to obtain the effective flip-angles at different slice-selective locations. Note that the rotation effect of the refocusing gradient should also be included in the slice profile computation, especially when the spin rephasing is not perfect, and the computed slice profile will be complex-valued. After computing the EPG signal evolution at all different $z$ locations, the transverse magnetization of the whole slice can be computed by summing up all zero-state signals $\tilde{F}_+(\mathbf{0}, z)$.

The slice profile computation by small tip-angle approximation (STA) is described by equation (2) as below,

$$SS(z) = M_{xy}^{RF}(z) = i\gamma M_0 \int_0^{T_{RF}} \bar{B}_{xy}(t) e^{-i\gamma z \int_t^{RF} G_z(s) ds} dt,$$

$$(2)$$

where $T_{RF}$ is the duration time of the RF pulse, $G_z(t)$ is amplitude of the linear gradient assuming $z$ is the slice selective direction, and $\bar{B}_{xy}(t)$ is the amplitude of the RF excitation pulse which is normalized to satisfy $i\gamma M_0 \int_0^{T_{RF}} \bar{B}_{xy}(t) dt = 1$. The effective flip-angle including slice profile correction at location $z$ then scales approximately linearly with the transmit RF field heterogeneity $B_1^+$, the slice profile $SS(z)$ and the amplitude of the RF pulse. The action of RF pulse with a given effective flip-angle in the EPG model can be computed by equation (10-12) in Reference[5].

Sources of error for standard EPG approximations could be caused by the following combined arguments. Firstly, one of the approximations involved in standard EPG is the fact that for SSFP trains, RF excitations are performed as instantaneous rotations. These effective instantaneous RF rotations angles are calculated either from STA (small tip-angle approximation as explained above[19,21]) or SLR (Shinnar-LeRoux[32]) methods, which are both approximation methods. Secondly, in SSFP sequences, the magnetization is repeatedly excited and thus what happens at a given excitation moment has effect on the subsequent time steps. As a consequence, effects from approximations in the RF excitation step cumulatively add up. Finally, the STA and SLR approximations are derived for initial condition equal to equilibrium. Even if in some cases these could be a good assumption (spoiling sequences "eliminate" transverse components) it is nonetheless an approximation, which can lead to cumulative errors in SSFP transient-state sequences.

A more accurate method to compute the magnetization response to an RF pulse is to discretize the RF pulse duration time $T_{RF}$ into $N_{RF}$ time intervals of length $\Delta t$, and numerically solve the Bloch equation for each time step by applying the operator splitting method[33,34]. Equation (1.58) in Reference[33] gives a first-order approximated solution for discretized time $t_{n+1}$,

$$\begin{bmatrix} M_x(t_{n+1}) \\ M_y(t_{n+1}) \\ M_z(t_{n+1}) \end{bmatrix} = \mathbf{R}_n \mathbf{D} \begin{bmatrix} M_x(t_n) \\ M_y(t_n) \\ M_z(t_n) \end{bmatrix} + (\mathbf{I} - \mathbf{D})\mathbf{M}_0,$$

(3)

Where $\mathbf{D} = \exp\left(\begin{bmatrix} e^{-\Delta t/T_2} & 0 & 0 \\ 0 & e^{-\Delta t/T_2} & 0 \\ 0 & 0 & e^{-\Delta t/T_1} \end{bmatrix}\right)$ is the relaxation operator, and $\mathbf{R}_n = \exp\left(\gamma\Delta t \begin{bmatrix} 0 & G_z(t_n) \cdot z & -B_1^+ \cdot B_y(t_n) \\ -G_z(t_n) \cdot z & 0 & B_1^+ \cdot B_x(t_n) \\ B_1^+ \cdot B_y(t_n) & -B_1^+ \cdot B_x(t_n) & 0 \end{bmatrix}\right)$ is the rotation operator, and $\mathbf{M}_0 = [0, 0, M_0]^\mathbf{T}$ is the magnetization in equilibrium state.

It is shown in Supplementary material (Appendix A) that any linear operator used in Bloch simulator can also be applied in the EPG model. Therefore, equation (3) for Bloch equation model can be transformed into the EPG model as below

$$\tilde{\mathbf{F}}(\mathbf{k}, t_{n+1}) = \begin{cases} \mathbf{SR}_n\mathbf{DS}^{-1}\tilde{\mathbf{F}}(\mathbf{k}, t_n) + \mathbf{S}(\mathbf{I} - \mathbf{D})\mathbf{M}_0, & \text{if } \mathbf{k} = 0, \\ \mathbf{SR}_n\mathbf{DS}^{-1}\tilde{\mathbf{F}}(\mathbf{k}, t_n), & \text{otherwise,} \end{cases}$$

(4)

where $\mathbf{S} = \begin{bmatrix} 1 & i & 0 \\ 1 & -i & 0 \\ 0 & 0 & 1 \end{bmatrix}$ is the similarity transformation matrix defined as in Reference[5]. We call the EPG model which computes the discretized RF excitation response by equation (4) EPG-Bloch, since equation (4) is derived from the approximate Bloch equation solution (3).

### 2.1.1 Computational Complexity

The EPG-Bloch model described above is more time-consuming comparing to the EPG model with slice profile correction, because it computes the RF pulse response in $N_{RF}$ time steps instead of one instantaneous rotation. When an MR sequence of $N_{TR}$ RF pulses is applied, $N_Z$ spatial points in the slice selective direction and $N_k$ configuration states in the EPG model are included, simulating the magnetization signal for one single voxel requires computing equation (4) for $N_{TR} \times N_Z \times N_k \times N_{RF}$ times. Taking $N_{TR} = 1000, N_Z = 32, N_k = 20$ and $N_{RF} = 100$ as an example, equation (4) needs to be computed by $6.4 \times 10^7$ times using the EPG-Bloch model. Note that even though the new EPG-Bloch model is computationally expensive, simulating gradient-spoiled sequences using the new EPG model is still more efficient comparing to multiple isochromat Bloch simulations. Theoretically, to fully simulate a gradient-spoiled sequence, one would need the same number of configuration states $N_k$ in EPG simulation as the number of isochromats $N_{iso}$ in the Bloch simulation, that is $N_k = N_{iso} = N_{TR}$.[35] However, in practice, since the configuration states in the EPG model decay through the time as a consequence of spin-spin and spin-lattice relaxation, $N_k$ can be substantially lower than $N_{TR}$, thus the number of configuration states required by the EPG model is much smaller than the number of spatial isochromats required by the Bloch simulator to achieve accurate simulation[23,35].

## 2.2 A Recurrent Neural Network as A Surrogate Model

As a consequence of the increased numerical complexity, the EPG-Bloch model requires a potentially prohibitively long computation time. In order to reduce the time needed to generate a large amount of magnetization signals, for example in MRF dictionary generation, an RNN network could be trained as a fast surrogate model to replace the new EPG-Bloch model simulation. RNN models can be effectively used to model time-dependent processes and especially ordinary differential equations, as shown in [25–27], therefore they are very suitable for MR signal computations. In contrast to previous work[13,14], we seek a unique surrogate model which can work for various sequence parameters, such as repetition time, number of RF excitations (sequence length) and flip-angle trains; therefore, retraining of the network will not be needed when we change the parameters of the sequences. Specifically, an RNN architecture with multiple stacked Gated Recurrent Units (GRU)[27,36] is selected; the RNN architecture used in this paper is shown in Figure 1. Figure 1(a) shows the RNN structure for the $n$-th time step, which includes three GRU layers and one Linear layer. The inputs for the $n$-th RF pulse include both tissue specific parameters $\boldsymbol{\theta}$, (e.g. $T_1$ and $T_2$), and time-dependent sequence parameters $\beta(n)$, such as $TR(n), TE(n)$ and flip-angle $\alpha(n)$.

All of the three GRU layers have the same structure, and they receive hidden state inputs $\boldsymbol{h}_1(n - 1), \boldsymbol{h}_2(n-1), \boldsymbol{h}_3(n-1)$ and return the updated $\boldsymbol{h}_1(n), \boldsymbol{h}_2(n), \boldsymbol{h}_3(n)$. The hidden states memorize information based on prior inputs, and can be interpreted as an alternative representation of spin states. The initial hidden states $\boldsymbol{h}_1(0), \boldsymbol{h}_2(0), \boldsymbol{h}_3(0)$ can be computed by adding an initial linear layer $Linear_{init}$, as shown in Figure 1(b). Input of the initial linear layer is the initial magnetization vector $\boldsymbol{M}_0$.

The hidden state output of the last GRU layer then passes to another linear layer, generating the output signal $M_{xy}(n)$, which is the transverse magnetization at the $n$-th echo time. Optionally, the linear layer could also directly compute additional derivative signal outputs, $dM_{xy}(n)/d\boldsymbol{\theta}$. Computing derivatives by a linear combination of the RNN outputs does not add computation time, therefore the proposed solution is a more efficient method compared to finite difference (FD) approximations or automatic differentiation (AD) methods.

Our RNN implementation of the EPG-Bloch model is available for download at https://gitlab.com/HannaLiu/rnn_epg.

# 3    Methods

## 3.1    Validation of the EPG-Bloch Model

To demonstrate the accuracy of the new proposed EPG-Bloch model, experiments were performed on a 1.5T clinical MR system (Ingenia, Philips Healthcare, Best, The Netherlands). A transient–state gradient-echo sequence with gradient spoiling and smoothly varying flip-angle train was used. Such transient-state pulse sequences have been used for quantitative MR experiments such as MRF[6] and MR-STAT[7,8]. A non-selective inversion pulse was applied at the beginning of the sequence to provide better $T_1$ encoding, and the waveform of the flip-angle train is shown in Figure 2 (a). A total number of 480 RF pulses was used in the sequence, each of which had a Gaussian-shaped waveform, a transverse slice thickness of 5 mm and a duration of 0.568 ms. Other settings were: inversion time $T_I = 7.74$ ms, repetition time $T_R = 7.38$ ms, and echo time $T_E = 3.73$ ms.

Three gel phantom tubes with different $T_1$ and $T_2$ values were imaged separately using the gradient-spoiled sequence described above, with the slice selective gradient aligned with the longitudinal axis of the tubes. In order to measure the transverse magnetization signals at each TR, the phase-encoding gradients were switched off; as a consequence, no spatial in-plane encoding was applied, since the material in the tube is homogeneous over the excited transverse slice. Transient state signals for the corresponding sequence were computed by both the conventional EPG model with small tip-angle approximation and the newly proposed EPG-Bloch model with RF pulse discretization. For both models, simulations were run for 32 sub-slices across a distance three times wider than the slice thickness to include the out-of-slice excitation, and 20 EPG configuration states were included. For the EPG-Bloch model, the Gaussian RF pulse was discretized into 16 equally-spaced time steps during the simulation. The measured data was fitted to the dictionary of simulated signals to find $T_1$ and $T_2$ values which gave the best match. Reference $T_1$ and $T_2$ values of the tubes were obtained from an interleaved inversion-recovery and multi spin-echo sequence (2DMix) provided by the MR vendor[37].

The code for both the conventional EPG and the proposed EPG-Bloch models were implemented in MATLAB based on existing code available online at https://web.stanford.edu/~bah/software/epg/. All EPG model simulations were run on an 8-core desktop PC with 3.7GHz CPUs.

## 3.2    Training and Validation of the RNN model

### Network structure specification

The RNN architecture described in previous Section 2.2 was selected for modeling gradient-spoiled sequence responses. Specifically, at the $n$-th time step, the inputs of the network were tissue parameters in logarithmic scale $\boldsymbol{\theta} = (\log T_1, \log T_2)^T$ and time-dependent sequence parameters $\boldsymbol{\beta} = (T_R(n), T_E(n), \alpha(n))^T$, and the outputs were $(M_{xy}(n), \partial M_{xy}(n)/\partial \boldsymbol{\theta})^T$, that was, both the magnetization and derivatives. At the first time step ($n = 0$), the input for the initial linear layer was the initial magnetization vector $\boldsymbol{M}_0 = (M_x(0), M_y(0), M_z(0))^T$. Each layer of the GRUs had 32 hidden states, and the whole network has in total 16643 trainable parameters.

**Dataset generation**

The training data were simulated from the new proposed EPG-Bloch model, containing a total number of 30000 magnetization signals. Each magnetization signal was simulated using a gradient-spoiled sequence with 1120 RF pulses, and all RF pulses had a Gaussian waveform shape with a duration of 1.0ms, and the same slice-selective and phase refocusing gradients leading to a slice thickness of 3 mm.

In the training dataset, each magnetization signal was computed using different input parameters. For tissue parameters $\theta = [\log T_1, \log T_2]^T$, $T_1$ and $T_2$ values were randomly sampled from logarithmic distributions ranging within $[0.1, 5]$ s and $[0.01, 2]$ s respectively. Only the parameter combinations with $T_1 \geq T_2$ were taken into account. For sequence parameters, $T_R$ and $T_E$ were chosen to be either time-constant or time-varying: for time-constant $T_R$ and $T_E$, one $T_R$ value was sampled uniformly within $[5, 20]$ ms for all RF pulses, and $T_E(n)$ value from $[0.3 * T_R, 0.7 * T_R]$ ms; for time-dependent $T_R$ and $T_E$, $T_R$ and $T_E$ values were randomly sampled for each of the RF pulses following the same sampling rule as the time-constant condition. For the flip-angle train $\alpha = [\alpha(1), \alpha(1), ..., \alpha(N_{TR})]^T$, every flip-angle $\alpha(n)$ was constrained to be in the interval $[0°, 120°]$. The flip-angle trains used for training the RNN network were all different and were randomly sampled from five different types of trains, including spline-interpolated functions with 5 control points (*Spline5*), spline-interpolated function with 11 control points (*Spline11*), sine-squared function with 5 sinusoidal lobes (*SinSquared5*), spline-interpolated function with a superimposed pseudo-random Gaussian component (*SplineNoise11*) and piecewise-constant functions (*PieceConstant5*). Details about how to generate these five different types of flip-angle trains are included in Supplementary material (Appendix B), and representative plots of the five types of waveforms are shown in Figure 3(a). In the dataset, each magnetization signal used a random flip-angle train generated from one of the five flip-angle train functions, so that the whole dataset had 6000 data signals generated using each type of the flip-angle trains for a total of 30000 unique trains. These five types of flip-angle trains were either smoothly varying functions (*Spline5*, *SinSquared5*, *Spline11*), or smoothly varying functions with random components (*SplineNoise11*), or piecewise-constant functions with random jumps (*PieceConstant5*). They provided various inputs feeding into the model such that various physical dynamics of the EPG model could be learned. The initial magnetization $M_0$ for the input of the initial linear layer was randomly chosen to be either $[0,0,-1]^T$ for sequence with an initial inversion pulse or $[0, 0, 1]^T$ for sequence without an inversion pulse.

To generate the training dataset, magnetization signals $M_{xy}$ were simulated by the new EPG-Bloch model using the generated input parameters described in the previous paragraph, and derivatives of the signals $\partial M_{xy} / \partial \theta$ were simulated by using automatic differentiation. In total 20000 data signals from the dataset were randomly selected for training the network coefficients, and the remaining 10000 signals were used for testing.

**Network training and validation**

The RNN network was built and trained using Tensorflow 2.2[38] on a Tesla V100 GPU with an Intel Xeon 2.6GHz processor. The Tesla V100 GPU has 32GB memory but we limited the maximum memory usage to be 12 GB. The training was run for 3000 epochs by an ADAM optimizer with adaptive learning rates[39], with a batch size of 200. An L1 loss function, Mean Absolute Error (MAE), was used during the training, and both MAE for the signal and its derivatives were weighted equally in the loss function.

The trained RNN model was subsequently used for predicting the magnetization and derivative signals with different tissue parameters and sequence parameters. The model was tested on the test data in the calculated dataset, and Normalized Root Mean Square Errors (NRMSEs) was used as the evaluation metric, and were calculated separately for signals and derivatives with different types of flip-angle trains.

**Runtime evaluation**

The trained RNN model can predict multiple signals in parallel in batches to accelerate the computation either on CPU or GPU. The type of hardware used for running the models is reported between brackets in the runtime evaluation experiment; the RNN model is always run on GPU without extra explanation for all other experiments. A maximum of batch size 6400 was selected in order to limit the memory usage to no more than 12GB. To evaluate the computational speed of the RNN (GPU) model prediction, different numbers of signals were predicted by the RNN model and the corresponding runtime was recorded. Every magnetization signal has the length of $N_{TR} = 1120$, and is simulated given the same sequence parameters as shown in Figure 4(a). When the number of magnetization signals $N_s$ was no larger than the maximum batch size 6400, we choose batch size $N_{batch} = N_s$; otherwise we set $N_{batch} = 6400$.

For performance comparison purposes, snapMRF (GPU)[11], the RNN (CPU) model and the proposed EPG-Bloch (CPU) model are all used for runtime tests. SnapMRF is an open source package for dictionary generation and signal matching in MRF. It allows for fast parallelizable GPU execution, and both physical models, Bloch equation and EPG model, are supported for signal simulation. To conform the snapMRF results equivalent to the EPG model with slice profile correction as described in Section 2.1, the snapMRF code needs to be repeated 32 times with different effective flip-angle trains. Note that snapMRF always treats RF excitation effects as instantaneous rotation, and therefore cannot be easily modified to realize EPG-Bloch model computations, thus we expect errors in the accuracy of the snapMRF model. Nonetheless, we decided to use snapMRF as it is publicly available and seems to be one of the best performing software packages in circulation.

Two different experiments were run for testing the runtime with respect to the number of signals. For fixed $B_1^+ = 1.0$, 6 different discretizations of the $T_1$ and $T_2$ domain were selected, resulting in a two-dimensional dictionary of $N_s = 100, 400, 800, 1600, 3200, 6400$ signals. For the case with various $B_1^+$ inputs, 4 different discretization were selected for $B_1^+, T_1$ and $T_2$ respectively, such that the number of different $B_1^+$ values would be 10, 20, 40 and 80, and total number of signals in the three-dimensional dictionary was $N_s = 10^3, 20^3, 40^3, 80^3$. For the varying $B_1^+$ case, only snapMRF (GPU) and RNN (GPU) were tested since the runtime for the other two CPU models would be prohibitively long. We point out that computation of the derivatives by the proposed RNN (GPU) model does not significantly increase the computation time since it requires only a small last linear step. For this reason, no explicit timing tests were run for the derivative parts. For the RNN (CPU/GPU) model, computations in the last linear layer (box on the top of Figure 1(a)) related to signal derivatives can be removed when no derivatives are required.

## 3.3   Applications of the RNN model

### 3.3.1   MRF reconstruction using RNN generated dictionary

**Numerical Brain Phantom**

To test the performance of the trained RNN model, the model was used for fast MRF dictionary generation. A gradient-spoiled transient-state sequence was used with the sequence parameters as shown in Figure 4(a). One radial k-space spoke with a golden-angle increment[40] was acquired for each TR with 224 sampling points along each readout spoke.

An MRF dictionary was generated by the surrogate RNN model with 100 logarithmically spaced $T_1$ values within [0.1,5] s, 100 logarithmically spaced $T_2$ values within [0.01,2] s and 40 uniformly spaced $B_1^+$ values within [0.8,1.2]. This resulted into a dictionary with 312480 atoms after the cases for which $T_2 > T_1$ were removed. For comparison, a second MRF dictionary was also generated by the EPG-Bloch model using the same reconstruction parameter values.

A numerical brain phantom[41] with a matrix size of 112×112 was used to simulate the synthetic data. The EPG-Bloch model was used to compute the magnetization signals for each voxel, and the acquired k-space data was then simulated by applying the non-uniform Fourier transform (NUFFT) to the volumetric magnetization signal. Complex Gaussian noise was added to the simulated k-space with a noise level SNR=40. The SNR was defined by the average k-space signal intensity divided by the standard deviation of the noise.

$T_1$, $T_2$, $B_1^+$ and proton density (PD) maps were reconstructed by the low-rank alternating direction method of multipliers (LR-ADMM) approach[42] using either the RNN generated or EPG-Bloch generated dictionary. Reconstruction parameters used in the LR-ADMM algorithm were: rank $R = 12$, ADMM penalty parameter $\mu = 0.015$, outer ADMM iterations = 20 and inner CG iterations = 10.

**In-vivo data**

In-vivo experimental data was collected on the 3.0 T MR system (Ingenia, Philips Healthcare) using the same sequence as in the previous numerical experiment. The same reconstruction parameters were used here, except that a homogenous $B_1^+ = 1.0$ was assumed and excluded from MRF reconstruction, and only 6 ADMM iterations were used here to avoid over-fitting during the LR-ADMM reconstructions.

### 3.3.2 Accelerating the optimal experimental design for MRF sequence

In statistical data analysis, the Cramér-Rao lower bound (CRLB) gives a lower bound on the variance of an unbiased estimator of a parameter[43], which is derived by inversion of the Fisher information matrix (FIM). The FIM is constructed by computing the derivatives of the signal with respect to the parameters to be estimated. The CRLB has been used for optimizing MRF sequence parameters for a small number of tissues with given $T_1$ and $T_2$ values[16,17,44]. For such an optimal experimental design problem, the objective function can be defined in several ways, depending on the criterion under consideration. A popular choice is to consider the weighted sum of the trace of the CRLB matrix. Optimizing CRLB objectives can be very computationally inefficient, because it requires computing a large number of magnetization signals and their derivatives, and it might even require the computation of the objective function's derivatives with respect to sequence parameters when using derivative-based optimization algorithms[45].

In this application, the RNN model was used for developing a computationally efficient method to optimize a gradient-spoiled MRF sequence. Given various sequence parameters, the RNN model is capable of computing large amounts of magnetization signals and derivative signals with respect to tissue parameters, therefore is very suitable for solving the CRLB optimization problem for optimal sequence design. For this aim, we departed from the standard derivative-based optimization algorithms, which are prone to find sub-optimal solutions given the non-convexity of the objective, and we chose to use a derivative-free, population-based optimization method[46] instead. In particular, we use the differential evolution (DE)[47,48], an algorithm which iteratively evolves the candidates in the population to obtain an improved solution on a derivative-free fashion. Note that here, by "derivative-free" we mean that the derivative of the objective function is not required during the optimization, however, in the CRLB-based optimal design problem, the objective function itself still requires derivative computation with respective to target tissue parameters. The Code for solving this optimal experimental design problem is implemented in Python using the differential evolution (DE) algorithm in Scipy and the trained RNN model.

Specifically, we conducted a simulation-based experiment to optimize the flip-angle train of an MRF sequence given two target tissues with $T_1/T_2 = 900/85$ ms and $T_1/T_2 = 500/65$ ms. For sequence parameters, we used time-constant $T_E/T_R = 4.9/8.7$ ms and $N_{TR} = 336$ (a different sequence length from the training data) with an initial inversion pulse, resulting into a very short acquisition time of 2.92s. The flip-angle train to be optimized was constrained to be only *Spline11* (See Appendix B) type waveform and the maximum flip-angle to be 90 degree. The constrained DE implementation provided by SciPy v1.5.0[49] was modified and used for solving the optimization problem. In each iteration of the modified DE algorithm, the RNN model was used once for computing the magnetization signals and derivatives for the whole population very fast. For the DE optimization, we set the population size to be equal to 10. Since *Spline11* requires 11 parameters to compute a complete flip-angle train and two target tissues are to be optimized, in each iteration a maximum of $10 \times 11 \times 2 = 220$ magnetization and derivative signals need to be computed. Other algorithm parameters using non-default values include: relative tolerance for convergence = 0.002, and maximum number of generations = 1000.

To evaluate the optimization results, two MRF reconstructions, with the original flip-angle train (made of the first 336 TRs shown in Figure 4(a)) and the optimized one, were conducted using the RNN generated dictionaries. The same numerical brain phantom, data generation method, and reconstruction algorithm as described in the previous Section 3.3.1 were used with only one main difference: in order to reduce the effects of k-space under-sampling, a spiral acquisition was used. The same spiral trajectory as in Reference[6,16] was used and one interleaf of a variable density spiral trajectory was acquired for each TR, whereas 48 interleaves were required for fully sampling the k-space. An ideal $B_1^+ = 1$ field was assumed everywhere and therefore it was not included in the MRF reconstructions.

# 4 Results

## 4.1 Validation of the new EPG model

Figure 2(b)-2(d) compare the experimental measured data with both the conventional EPG and the proposed EPG-Bloch model predictions. For larger flip-angle segments (around the $336^{th}$ TR index), as shown in the right column of magnified figures, simulated signals by the conventional EPG model are significantly larger than the measured data, while the new EPG-Bloch model signals show an improved match.

The reconstructed $T_1$ and $T_2$ values by using both EPG models are summarized in Table 1. For each of the three gel tubes, the fitted $T_2$ values by the new EPG-Bloch model are all slightly higher (about 10 ms) than the fitted $T_2$ values by the conventional EPG model, and are closer to the measured reference values by the reference sequence. The fitted $T_1$ values by both models show relatively small differences, and are both in good agreement with measured reference values. In this experiment, the better accuracy of the new proposed EPG-Bloch model is proved, motivating the usage of this accurate but computationally more expensive EPG- Bloch model for quantitative MR methods.

## 4.2 Validation of the RNN model

**Network validation results**

The total training time was approximately 8 hours, and the overall results for the RNN model validation are summarized in Table 1. The NRMSEs for five different types of flip-angle trains are given for the signals and derivatives respectively. It shows that the RNN results agree well with the EPG-Bloch results. The RNN results are less accurate for *PieceConstant5* flip-angle trains comparing to the other types. This problem may be solved by increasing the portion of the *PieceConstant5* type data used for training. Since this type of train is not common in transient state acquisitions, we did not perform additional trainings.

Two examples of sample magnetization and derivative signals generated from RNN are shown in Figure 4 and Figure 5. Figure 4(a) gives the sequence parameters used for a flip-angle train which belongs to the *SinSquared* type. Figure 4(b) shows RNN generated magnetization and derivatives for different $T_1$ and $T_2$ values. The three combinations of $T_1$ and $T_2$ values are chosen to be close to white matter, gray matter and cerebrospinal fluid (CSF) tissue parameters. All these RNN results match well to the EPG-Bloch results. Figure 5 shows similar results as in Figure 4. However, the sequence used here is longer than the ones used for training, and the flip-angle train is a combination of *Spline11* and *PieceConstant5* type, which is not included in the training dataset. The RNN results still match well to the EPG-Bloch results, showing good generalization properties of the trained RNN model.

**Runtime evaluation**

Figure 6 shows the runtime comparison results for the RNN (GPU) and snapMRF (GPU). As shown in both Figure 6(a) and 6(b), all the runtime curves grow approximately linearly with respect to the number of signals. For the fixed $B_1^+$ condition in Figure 6(a), RNN (GPU) requires ~100 times less runtime than snapMRF (GPU) for a dataset with 6400 signals, and ~200 times less runtime than the RNN (CPU) model and ~34000 times less than the EPG-Bloch (CPU) model. For the various $B_1^+$ condition in Figure 6(b), RNN (GPU) outperforms snapMRF (GPU) by a factor of 68 for a large dataset with 512000 signals.

The runtime comparison is divided into fixed $B_1^+$ condition and various $B_1^+$ condition, because the kernels in snapMRF are not parallelized for data points with different $B_1^+$ values. For example, snapMRF takes 144 seconds for a dataset of 6400 signals with 20 different $B_1^+$ values, but only takes 15.7 seconds for the same size of dataset with one fixed $B_1^+$ value. However for the RNN (CPU/GPU) model, $B_1^+$ values and tissue parameter inputs do not affect the runtime. The RNN runtime is only affected by the number of the signals and the sequence length $N_{TR}$. Note that the acceleration rate of the RNN (GPU) model comparing to snapMRF (GPU) slightly decreases for larger number of signals, but RNN (GPU) model still performs much faster, because of the repetitive computations required by the slice-profile correction in snapMRF.

### 4.3 Applications of the RNN-EPG model

#### 4.3.1 MRF reconstruction using an RNN-EPG generated dictionary

**Numerical Brain Phantom**

MRF reconstruction results using dictionaries generated by different models are shown in Figure 7. Mean relative errors are shown in the upper right corner of each relative error maps. Using the EPG-Bloch model and RNN model generated dictionaries, the reconstructions show high agreement with the ground truth maps, with the surrogate RNN model results having slightly higher relative errors compared to the EPG-Bloch model. In contrast, using the conventional EPG model generated dictionary, the $T_2$, $B_1^+$ and $PD$ maps are poorly reconstructed: overall, underestimated $T_2$ values are observed, and the $B_1^+$ map reconstruction fails.

For this numerical experiments, three reconstruction parameters, $T_1$, $T_2$ and $B_1^+$, are included, resulting in a dictionary with 312480 atoms. Generating the whole dictionary using the EPG-Bloch model on CPU takes about 71 hours, however the dictionary generation using the RNN model on GPU takes less than 10 seconds. This suggests the extreme efficiency of using the RNN model for MRF dictionary generation with negligible loss in image quality.

**In-vivo data**

MRF reconstruction results using in-vivo data are shown in Figure 8. Mean absolute differences between EPG-Bloch reconstruction and RNN reconstruction are 1.74%, 3.46% and 2.02% for $T_1$, $T_2$ and abs($PD$) maps respectively, and relatively larger differences mostly exist in CSF regions.

#### 4.3.2 Accelerating the optimal experimental design for MRF sequence

The flip-angle train obtained after solving the optimal experimental design problem described in Section 3.3.2 is shown in Figure 9(a). Compared to previously reported results[16,17], the optimized flip-angle train shows very similar smoothened trapezoidal pattern. Solving this optimization problem requires about 580 iterations to converge using the DE algorithm for a total of approximately 70000 signal derivatives computations by the RNN model. The total runtime is 41 seconds when using the RNN model. In previous works[16,17], similar optimal experimental design problems require at least one CPU hour to solve. Our proposed implementation using the RNN model shows a significantly reduced runtime by approximately two orders of magnitude.

Figure 9(b) shows the MRF-reconstructed $T_1$, $T_2$ and $PD$ maps using the original and optimized flip-angle trains. It can be seen that the optimized sequence improves the accuracy of all the three reconstructed maps comparing to the original sequence, and the improvement in $T_2$ maps is the most significant.

## 5 Discussion

In this work, we have presented a new EPG-Bloch model which accurately models the RF excitation effects. For the EPG-Bloch model, rotation operations are applied in the configuration state domain to compute RF pulse excitations by sequentially discretized sub-steps, similar as computing RF pulse excitation effects in spin domain using the Bloch equation. The effects of other components of the sequence, e.g. the spoiler gradients, are simulated in the configuration state domain. By applying the new EPG-Bloch model, more accurate magnetization signals can be computed by tracing the signal evolution of just 15~20 configurations states. In comparison, the conventional EPG model using the slice-profile correction method is less accurate, especially when larger flip-angles are employed. A full Bloch equation model is less efficient for computing the spoiler gradient responses, requiring simulation of hundreds to thousands of spins. In conclusion, we combined the best ingredients of both model (Bloch and EPG) into a combined model which is faster than the Bloch simulation and more accurate than the conventional EPG simulator.

Recent work[24] has also incorporated the time-dependent RF waveform response for accurate modeling of the slice profile effects (the ssEPG model). In ssEPG, the RF excitation is simulated in the configuration space (frequency-domain) along the slice-selective dimension, whereas in our EPG-Bloch model we propose to simulate the RF excitation separately for each sub-slice (space-domain). Both models seem to agree very well with experimental measurements and to be more accurate alternatives to the standard EPG. Extended comparison of the two methods would go beyond the scope of this work and is left to future work. It should be noticed that, similar to EPG-Bloch, also the ssEPG model is computationally slower than conventional EPG, indicating the necessity for acceleration. The latter is an additional motivation for the main contribution of this paper, that is, acceleration by RNN surrogates.

In order to reduce the runtime for signal computations using the new EPG-Bloch model, we have proposed an RNN model as a surrogate for fast signal computations. Specifically, we have trained the network to learn magnetization signals and derivatives for a gradient-spoiled sequence, with arbitrary tissue parameters $T_1, T_2$, and sequence parameters such as sequence length, repetition time, echo time, flip-angle amplitudes and initial magnetization as inputs of the model. The trained RNN model computes both the magnetization signals and $T_1$ and $T_2$ derivatives with high accuracy, and is about 36000 times faster than the EPG-Bloch model and between one and three orders of magnitude faster than the snapMRF package. Two use cases are provided, namely MRF dictionary generation and optimal experimental design, indicating the broad application prospects of the RNN model.

One main advantage of the RNN model is that it is capable of learning MR signals for various sequence parameters, such as flip-angle train, repetition time and sequence length. After training with a relatively small dataset with 20000 MR signals, the RNN model could compute magnetization signals for sequences with new sequence parameters without the need of retraining. This suggests the RNN model could be used for fast signal computations when sequences need to be modified. The RNN model can also be used for accelerating the sequence parameter optimization process, when signal and derivatives for different sequence parameters need to be computed. Our numerical experiment results show that optimizing a flip-angle train of length 400 for two target tissues requires only 41 seconds when running on GPU, and may be further used for accelerating more complicated sequence optimization problems in the future, for example, optimizing sequences for more reconstruction parameters such as magnetization transfer[50] or $B_1^+$.

The main purpose of using the RNN model is to accelerate large-scale MR signal computations. When running on GPU, the proposed RNN model requires at most 10 seconds for generating a large MRF dictionary with $2 * 10^5$ magnetization signals. Training the RNN requires a relatively small training dataset with $2 * 10^4$ magnetization signals, only 10% size of the dictionary, and takes a relatively long training time, in our experiment more than 8 hours for 3000 epochs. Generation of the training dataset by the EPG-Bloch model also adds about another 13 hours of computation time. However, dataset generation and network training need to be done only once, the trained network requires very little storage space, and can be repeatedly loaded and used for signal computations with various tissue parameters and sequence parameters each time.

Another advantage of the proposed RNN model is that it can compute both the magnetization signal and the signal derivatives with respect to the parameters of interest ($T_1$ and $T_2$ in current experiments) simultaneously. For the RNN model, derivatives can also be computed by using either finite difference method or automatic differentiation; however, the runtime would be at least doubled to include also the derivative computation. In contrast, we train our RNN model to compute both signal and derivative at the same time, such that the hidden states of the RNN units contain the derivative information, and the derivatives can be computed by a weighted sum of the hidden states. Our proposed RNN model requires negligible additional runtime for derivative computation.

The current RNN model learns the signal evolution for a gradient-spoiled sequence with a Gaussian RF excitation pulse. Since RNN architecture is very effective for learning various time-dependent processes, especially those which can be exactly modelled by ordinary differential equations, we believe the proposed RNN model is also able to learn different types of sequences, such as gradient-spoiled sequence with a different RF shape, RF spoiled sequences or balanced sequences, or different physical models, such as presented in Reference[24]. Re-training for learning new types of sequences or other physical models will be required, but future training can be accelerated by transfer learning approaches[51] in the machine learning field, in which knowledge learned from previous trainings can be applied to a new but related problem.

The current trained RNN model can compute MR signals for various sequence parameter inputs and initial conditions. For flip-angle train inputs, five different types of waveforms were used, including all commonly used patterns for qMRI applications, and many flip-angle trains outside these five types can still be accurately computed. For repetition time inputs, $T_R$ can be either constant or time-varying in a random fashion throughout the sequence. For initial magnetization, two different initial values, either $[0, 0, -1]^T$ for perfect inversion pulse at the beginning or $[0, 0, 1]^T$ for no inversion pulse was used. In the future, in order to train an RNN model for more sequence parameter and initial magnetization inputs, RNNs with more stacked layers or more hidden states may be required. Adding stacked units or hidden states in the RNN model makes it possible to learn more sophisticated physical model responses, but also requires longer runtime for both training and signal computation, thus the trade-off between the model expressivity and computation time needs to be considered when applying new RNN architecture.

Although the RNN model is computationally very efficient, it is still slower than other existing deep learning approaches. For example, it is reported in Reference[14] that generating an MRF dictionary of size 1000 by 5930 takes 0.3 seconds on CPU. Generating an MRF dictionary of the same size using the RNN model takes 7.2 seconds on a similar CPU. Recurrent neural networks are relatively slow comparing to other network architectures, because they sequentially compute the signal for each time step. Further acceleration of the RNN model may be achieved by using more powerful computer capacity and state-of-art RNN inference libraries[52].

# 6   Conclusion

This work proposed a new EPG-Bloch model for simulating transient-state gradient-spoiled MR sequences, and trained an RNN as a fast surrogate of the EPG-Bloch model for large-scale signal computations. By comparing to the measured phantom data, we showed that the proposed EPG-Bloch model is more accurate than the standard EPG with small tip-angle approximation for imperfect slice profile correction. We further demonstrated that the RNN (EPG) model is computationally very efficient, at least four orders of magnitude faster than the EPG-Bloch simulation, and between one and three orders of magnitude faster than the GPU-accelerated EPG package snapMRF. Example experiments demonstrated that the RNN model can be efficiently used for computing large-scale ($>10^5$) MR signal dictionaries and derivatives for different qMRI applications such as MRF within tens of seconds.

# Acknowledgements

First author receives a scholarship granted by China Scholarship Council (CSC, #201807720088). The authors are grateful to Mr. Tom Bruijnen for the fruitful discussions.

# References

1. Filo S, Shtangel O, Salamon N, et al. Disentangling molecular alterations from water-content changes in the aging human brain using quantitative MRI. *Nat Commun*. 2019;10(1):1-16.

2. Piredda GF, Hilbert T, Granziera C, et al. Quantitative brain relaxation atlases for personalized detection and characterization of brain pathology. *Magn Reson Med*. 2020;83(1):337-351.

3. Ma D, Gulani V, Seiberlich N, et al. Magnetic resonance fingerprinting. *Nature*. 2013;495(7440):187-192.

4. Sbrizzi A, van der Heide O, Cloos M, et al. Fast quantitative MRI as a nonlinear tomography problem. *Magn Reson Imaging*. 2018;46:56-63.

5. Weigel M. Extended phase graphs: dephasing, RF pulses, and echoes-pure and simple. *J Magn Reson Imaging*. 2015;41(2):266-295.

6. Jiang Y, Ma D, Seiberlich N, Gulani V, Griswold MA. MR fingerprinting using fast imaging with steady state precession (FISP) with spiral readout. *Magn Reson Med*. 2015;74(6):1621-1631.

7. van der Heide O, Sbrizzi A, Bruijnen T, van den Berg CAT. Extension of MR-STAT to non-Cartesian and gradient-spoiled sequences. In: *Proceedings of the 2020 Virtual Meeting of the ISMRM*. 2020:0886.

8. Mandija S, D'Agata F, Liu H, et al. A five-minute multi-parametric high-resolution whole-brain MR-STAT exam: first results from a clinical trial. In: *Proceedings of the 2020 Virtual Meeting of the ISMRM*. 2020:0558.

9. Davies M, Puy G, Vandergheynst P, Wiaux Y. A compressed sensing framework for magnetic resonance fingerprinting. *SIAM J Imaging Sci*. 2014;7(4):2623-2656.

10. Körzdörfer G, Jiang Y, Speier P, et al. Magnetic resonance field fingerprinting. *Magn Reson Med*. 2019;81(4):2347-2359.

11. Wang D, Ostenson J, Smith DS. snapMRF: GPU-accelerated magnetic resonance fingerprinting dictionary generation and matching using extended phase graphs. *Magn Reson Imaging*. 2020;66:248-256.

12. Xanthis CG, Aletras AH. coreMRI: A high-performance, publicly available MR simulation platform on the cloud. *PLoS One*. 2019;14(5):e0216594.

13. Yang M, Jiang Y, Ma D, Mehta BB, Griswold MA. Game of learning Bloch equation simulations for MR fingerprinting. *arXiv Prepr arXiv200402270*. 2020.

14. Hamilton JI, Currey D, Griswold M, Seiberlich N. A neural network for rapid generation of cardiac MR fingerprinting dictionaries with arbitrary heart rhythms. In: *Proceedings of the 27th Annual Meeting of ISMRM,Montreal*. 2019:2421.

15. Sbrizzi A, Bruijnen T, van der Heide O, Luijten P, van den Berg CAT. Dictionary-free MR Fingerprinting reconstruction of balanced-GRE sequences. *arXiv Prepr arXiv171108905*. 2017.

16. Zhao B, Haldar JP, Liao C, et al. Optimal experiment design for magnetic resonance fingerprinting:

Cramer-Rao bound meets spin dynamics. *IEEE Trans Med Imaging*. 2018;38(3):844-861.

17. Lee PK, Watkins LE, Anderson TI, Buonincontri G, Hargreaves BA. Flexible and efficient optimization of quantitative sequences using automatic differentiation of Bloch simulations. *Magn Reson Med*. 2019;82(4):1438-1451.

18. van der Heide O, Sbrizzi A, Luijten PR, van den Berg CAT. High-resolution in vivo MR-STAT using a matrix-free and parallelized reconstruction algorithm. *NMR Biomed*. 2020;33(4):e4251.

19. da Cruz G, Bustin A, Jaubert O, Schneider T, Botnar RM, Prieto C. Sparsity and locally low rank regularization for MR fingerprinting. *Magn Reson Med*. 2019;81(6):3530-3543.

20. Emmerich J, Flassbeck S, Schmidt S, Bachert P, Ladd ME, Straub S. Rapid and accurate dictionary-based T2 mapping from multi-echo turbo spin echo data at 7 Tesla. *J Magn Reson Imaging*. 2019;49(5):1253-1262.

21. Hilbert T, Xia D, Block KT, et al. Magnetization transfer in magnetic resonance fingerprinting. *Magn Reson Med*. 2020;84(1):128-141.

22. Ma D, Coppo S, Chen Y, et al. Slice profile and B1 corrections in 2D magnetic resonance fingerprinting. *Magn Reson Med*. 2017;78(5):1781-1789.

23. Ben-Eliezer N, Sodickson DK, Block KT. Rapid and accurate T2 mapping from multi--spin-echo data using Bloch-simulation-based reconstruction. *Magn Reson Med*. 2015;73(2):809-817.

24. Ostenson J, Smith DS, Does MD, Damon BM. Slice-selective extended phase graphs in gradient-crushed, transient-state free precession sequences: An application to MR fingerprinting. *Magn Reson Med*. 2020.

25. Niu MY, Horesh L, Chuang I. Recurrent neural networks in the eye of differential equations. *arXiv Prepr arXiv190412933*. 2019.

26. Sundermeyer M, Schlüter R, Ney H. LSTM neural networks for language modeling. In: *Thirteenth Annual Conference of the International Speech Communication Association*. ; 2012.

27. Cho K, Van Merriënboer B, Gulcehre C, et al. Learning phrase representations using RNN encoder-decoder for statistical machine translation. *arXiv Prepr arXiv14061078*. 2014.

28. Banerjee K, Georganas E, Kalamkar DD, et al. Optimizing deep learning rnn topologies on intel architecture. *Supercomput Front Innov*. 2019;6(3):64-85.

29. Chetlur S, Woolley C, Vandermersch P, et al. cudnn: Efficient primitives for deep learning. *arXiv Prepr arXiv14100759*. 2014.

30. Appleyard J, Kocisky T, Blunsom P. Optimizing performance of recurrent neural networks on gpus. *arXiv Prepr arXiv160401946*. 2016.

31. Pauly J, Nishimura D, Macovski A. A k-space analysis of small-tip-angle excitation. *J Magn Reson*. 1989;81(1):43-56.

32. Buonincontri G, Sawiak SJ. MR fingerprinting with simultaneous B1 estimation. *Magn Reson Med*. 2016;76(4):1127-1135.

33. Valenberg W van. Radiofrequency pulse design through optimal control and model order reduction of the Bloch equation. 2015.

34. Glowinski R, Osher SJ, Yin W. *Splitting Methods in Communication, Imaging, Science, and Engineering*. Springer; 2017.

35. Malik SJ, Sbrizzi A, Hoogduin H, Hajnal J V. Equivalence of EPG and isochromat-based simulation of MR signals. In: *In Proceedings of the 24th Annual Meeting of ISMRM, Singapore*. 2016:3196.

36. Hermans M, Schrauwen B. Training and analysing deep recurrent neural networks. In: *Advances in Neural Information Processing Systems*. ; 2013:190-198.

37. In den KJJ, Cuppen JJ. RLSQ: T1, T2, and rho calculations, combining ratios and least squares. *Magn Reson Med*. 1987;5(6):513.

38. Abadi M, Agarwal A, Barham P, et al. Tensorflow: Large-scale machine learning on heterogeneous distributed systems. *arXiv Prepr arXiv160304467*. 2016.

39. Kingma DP, Ba J. Adam: A method for stochastic optimization. *arXiv Prepr arXiv14126980*. 2014.

40. Winkelmann S, Schaeffter T, Koehler T, Eggers H, Doessel O. An optimal radial profile order based on the Golden Ratio for time-resolved MRI. *IEEE Trans Med Imaging*. 2006;26(1):68-76.

41. Aubert-Broche B, Evans AC, Collins L. A new improved version of the realistic digital brain phantom. *Neuroimage*. 2006;32(1):138-145.

42. Assländer J, Cloos MA, Knoll F, Sodickson DK, Hennig J, Lattanzi R. Low rank alternating direction method of multipliers reconstruction for MR fingerprinting. *Magn Reson Med*. 2018;79(1):83-96.

43. Cowan G. *Statistical Data Analysis*. Oxford university press; 1998.

44. van Riel MHC, Yu Z, Hodono S, et al. Optimization of MR Fingerprinting for Free-Breathing Quantitative Abdominal Imaging. *arXiv Prepr arXiv200602928*. 2020.

45. Nocedal J, Wright S. *Numerical Optimization*. Springer Science & Business Media; 2006.

46. Kara D, Fan M, Hamilton J, Griswold M, Seiberlich N, Brown R. Parameter map error due to normal noise and aliasing artifacts in MR fingerprinting. *Magn Reson Med*. 2019;81(5):3108-3123.

47. Storn R, Price K. Differential evolution--a simple and efficient heuristic for global optimization over continuous spaces. *J Glob Optim*. 1997;11(4):341-359.

48. Lampinen J. A constraint handling approach for the differential evolution algorithm. In: *Proceedings of the 2002 Congress on Evolutionary Computation. CEC'02 (Cat. No. 02TH8600)*. Vol 2. ; 2002:1468-1473.

49. Virtanen P, Gommers R, Oliphant TE, et al. SciPy 1.0: fundamental algorithms for scientific computing in Python. *Nat Methods*. 2020;17(3):261-272.

50. Malik SJ, Teixeira RPAG, Hajnal J V. Extended phase graph formalism for systems with magnetization transfer and exchange. *Magn Reson Med*. 2018;80(2):767-779.

51. Tan C, Sun F, Kong T, Zhang W, Yang C, Liu C. A survey on deep transfer learning. In: *International Conference on Artificial Neural Networks*. ; 2018:270-279.

52. Holmes C, Mawhirter D, He Y, Yan F, Wu B. Grnn: Low-latency and scalable rnn inference on gpus. In: *Proceedings of the Fourteenth EuroSys Conference 2019*. ; 2019:1-16.

**Figure 1.** RNN structure for learning the EPG model. (a) RNN architecture with three stacked Gated Recurrent Units (GRU) for the $n$-th time step. At each time step, GRU1 receives inputs x(n) including tissue parameter $\theta$ and time-varying sequence parameter $\beta(n)$. The hidden states $\boldsymbol{h}_1(n), \boldsymbol{h}_2(n), \boldsymbol{h}_3(n)$ are computed and used for the next time step. A Linear layer is added after GRU3 to compute the magnetization and derivatives using the hidden states $\boldsymbol{h}_3(n)$. (b) An initial linear layer, $Linear_{init}$, is used for computing the initial hidden state $\boldsymbol{h}_1(0), \boldsymbol{h}_2(0), \boldsymbol{h}_3(0)$ from initial magnetization $\boldsymbol{M}_0$.

(a)

At the $n$-th RF pulse ($0 < n \leq N_{TR}$):

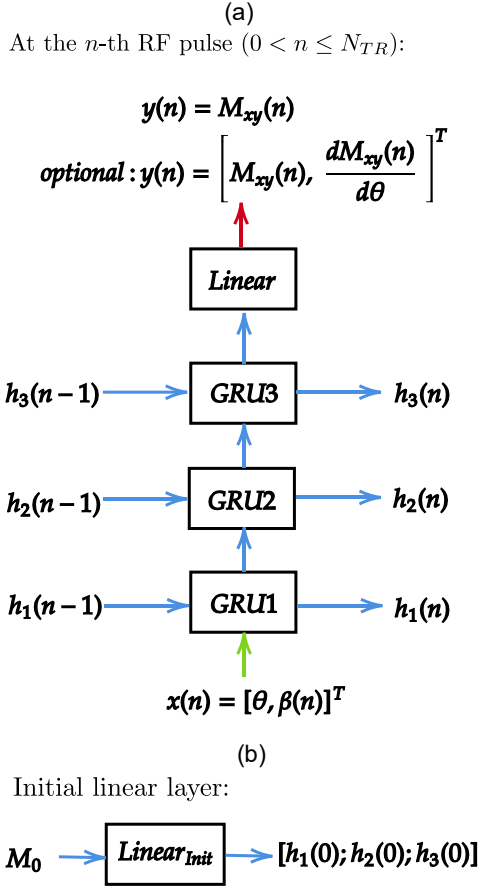

(b)

Initial linear layer:

$M_0 \rightarrow \boxed{Linear_{init}} \rightarrow [h_1(0); h_2(0); h_3(0)]$

**Figure 2.** Experimental validation of the EPG-Bloch model. (a) Flip-angle train for the transient-state gradient-spoiled sequence. (b-d) Experimental data compared with both conventional EPG and EPG-Bloch results for the three tubes with different $T_1$ and $T_2$ values. A magnified portion is shown on the right of each plot. (e) Fitted $T_1$ and $T_2$ values using conventional EPG and EPG-Bloch generated dictionaries, comparing with 2DMix reference results.

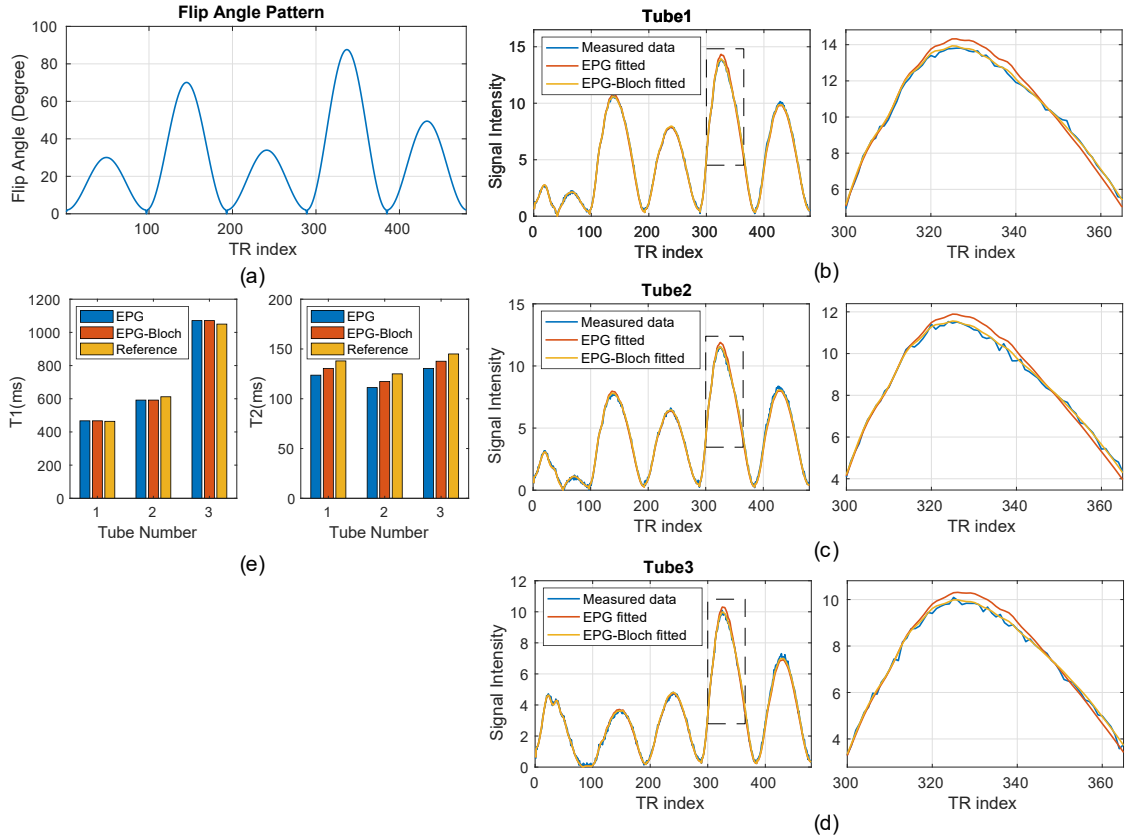

**Figure 3**. Example flip-angle trains for training the RNN-EPG model. Five flip-angle trains sampled from each different type of train functions are plotted.

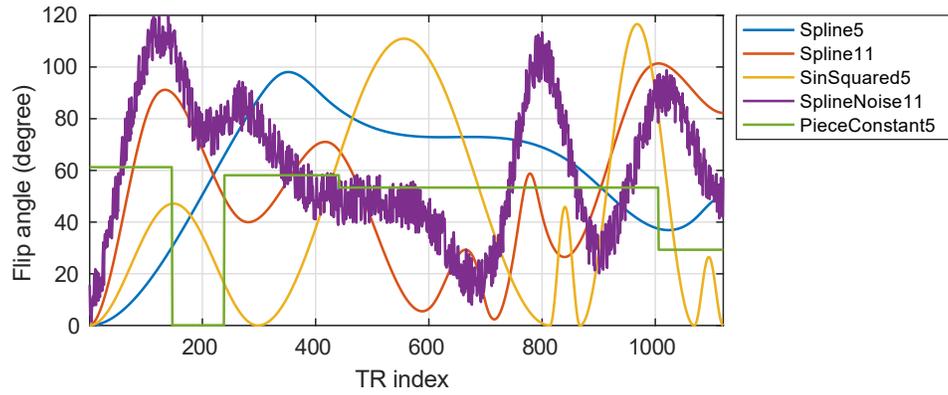

**Table 1.** Validation of the RNN-EPG model for different types of flip-angle trains. NRMSE (Normalized Root Mean Square Error) are computed for the validation dataset. Errors are computed respectively for data with different types of flip-angle trains, and signal and derivative errors are computed separately.

|  | *Spline5* | *Spline11* | *SinSquared5* | *SplineNoise11* | *PieceConstant5* |
|---|---|---|---|---|---|
| Signal | 0.418% | 0.574% | 0.780% | 1.285% | 5.155% |
| Derivative | 0.801% | 1.008% | 1.694% | 2.037% | 3.913% |

**Figure 4.** Sample magnetization and derivative signals generated with the surrogate RNN model. (a) Sequence parameter (flip-angle train, time-constant TE and TR) plots. Initial magnetization is $M_0 = [0,0,-1]^T$ after the inversion pulse. (b) Magnetization signal and derivative plots compared with EPG-Bloch results given different $T_1$ and $T_2$ values.

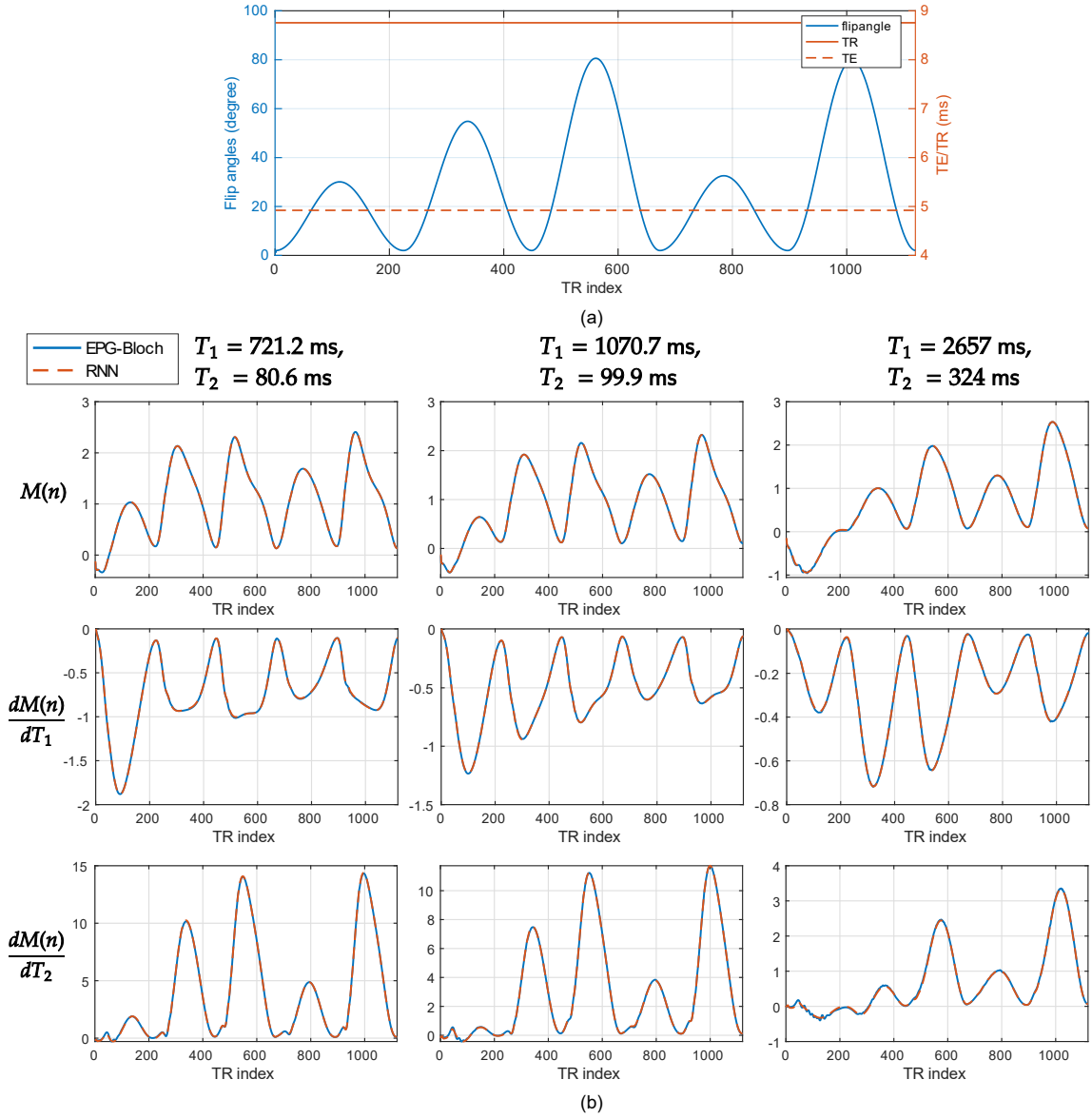

(a)

(b)

**Figure 5.** Another sample magnetization and derivative signals generated with the surrogate RNN model. (a) Sequence parameter (flip-angle train, time-varying TE and TR) plots. No inversion pulse is applied. (b) Magnetization signal and derivative plots compared with EPG-Bloch results given different $T_1$ and $T_2$ values.

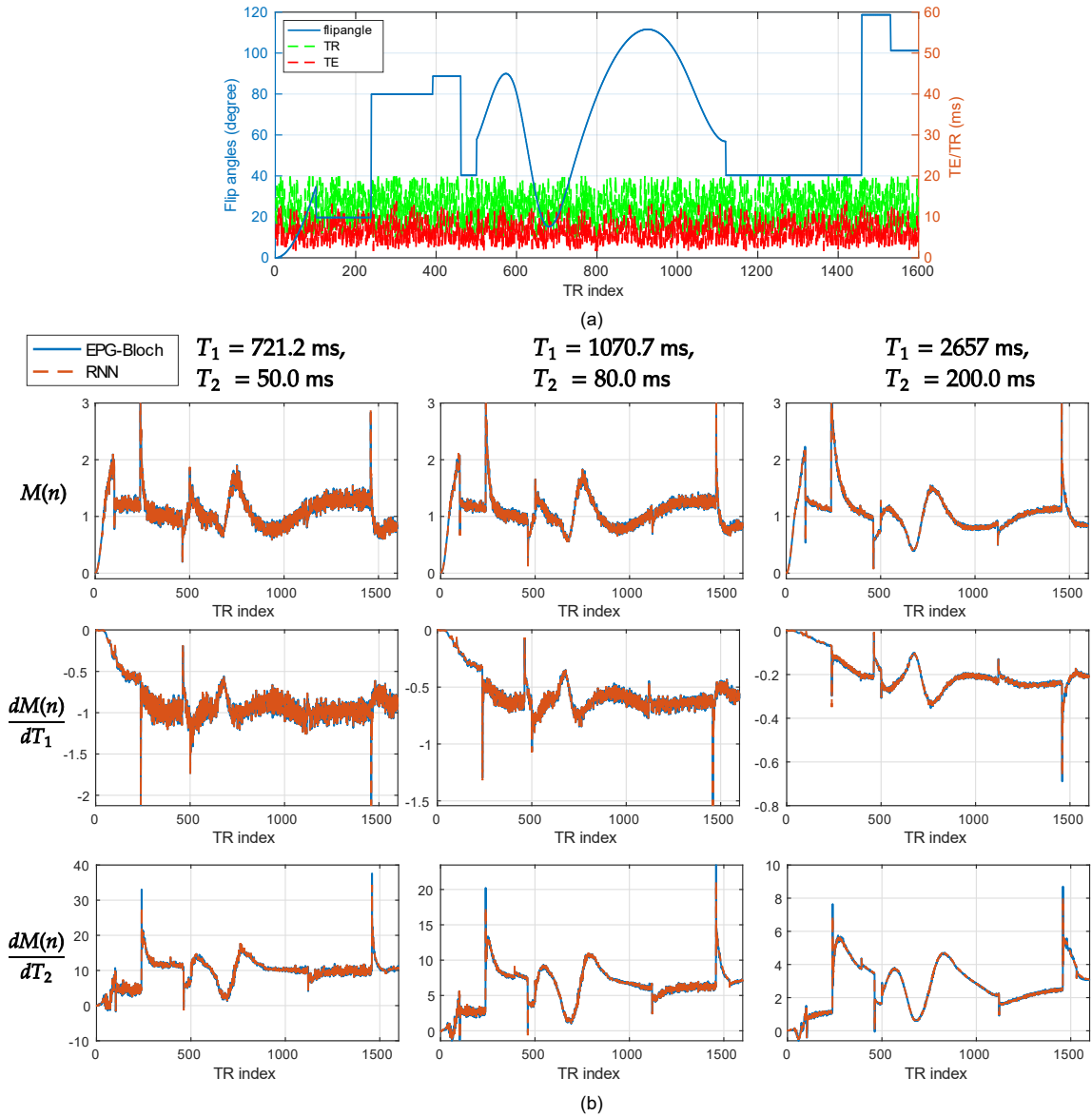

(a)

(b)

**Figure 6.** RNN(GPU) runtime comparison with snapMRF(GPU), RNN(CPU) and EPG(CPU). (a) Runtime comparison for the two-dimensional $(T_1, T_2)$ dictionary generation. (b) Runtime comparison for the three-dimensional $(T_1, T_2, B_1^+)$ dictionary.

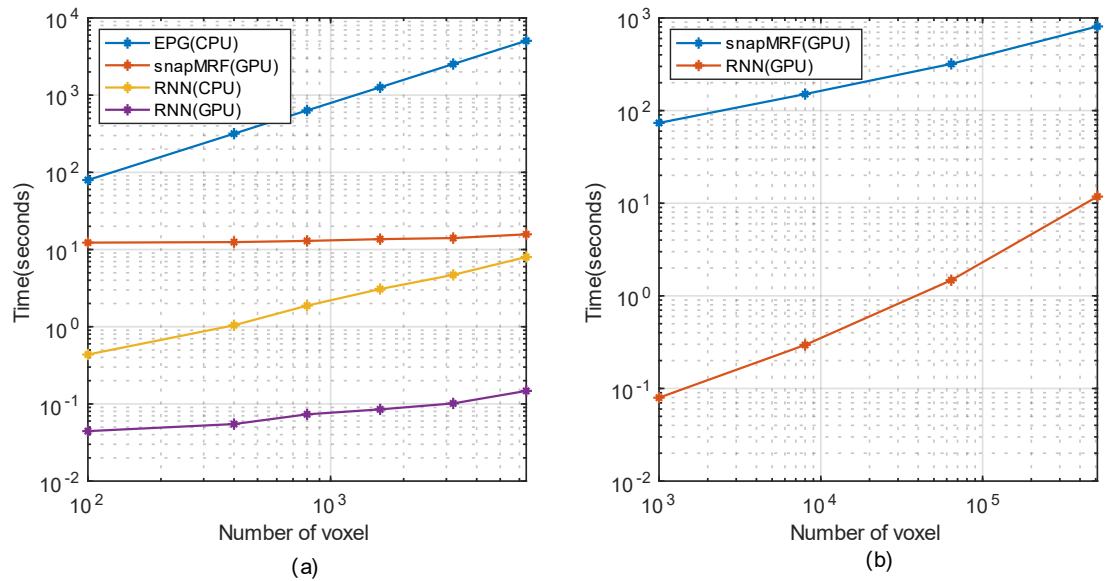

(a)                                                    (b)

**Figure 7.** MR Fingerprinting reconstructions of the numerical brain phantom for different signal models. Mean absolute percentage error (MAPE) values are reported on the error maps.

[First column] Ground truth $T_1$, $T_2$, $B_1^+$ and $PD$ maps for the numerical brain phantom. [Second, third and fourth columns] Reconstructed maps and absolute relative error maps for MRF using different dictionaires.

[Second column]: MRF dictionary generated by original EPG model with small tip-angle approximation;

[Third column]: MRF dictionary generated by the new EPG-Bloch model;

[Fourth column]: MRF dictionary generated by the RNN model;

Both the EPG-Bloch and the RNN reconstruction results show great agreements with the ground truth maps, with RNN results having slightly higher relative errors, whereas the MRF(EPG) results have relatively high reconstruction errors, especially for $T_2$, $B_1^+$ and $PD$ maps.

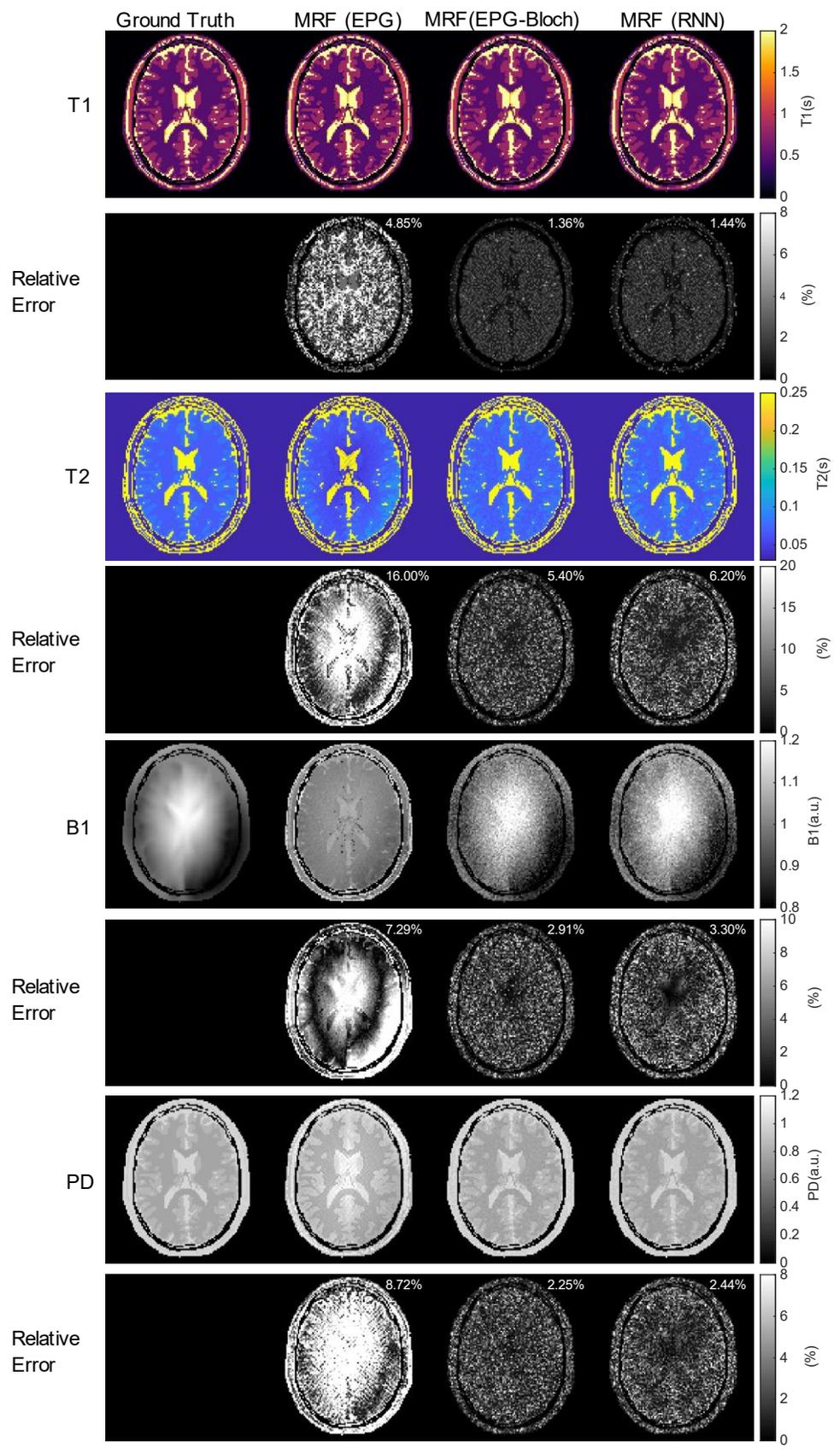

**Figure 8.** MRF reconstructions of in-vivo data using EPG-Bloch and RNN generated dictionaries.

[First, second and third rows] $T_1$, $T_2$, and $PD$ maps for the in-vivo brain data.

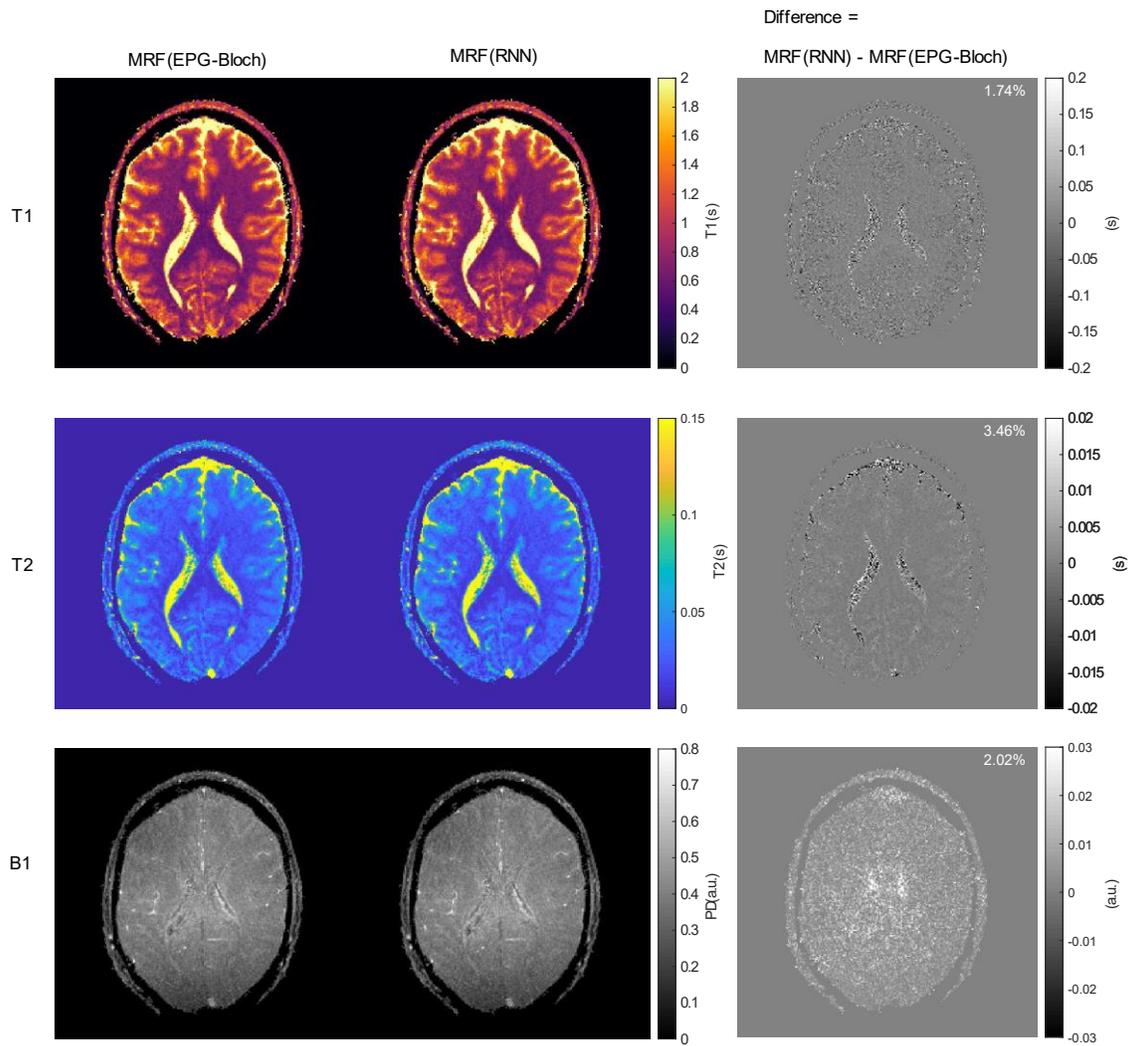

**Figure 9.** Optimal numerical experimental design results. (a) The original and optimized flip-angle trains. (b) MRF reconstruction results using the two different flip-angle trains. First column: Ground truth $T_1$, $T_2$ and $PD$ maps for the numerical brain phantom. Second and third columns: Reconstructed MRF maps and absolute relative error maps obtained using, respectively, the original and the optimized flip-angle trains. Mean absolute percentage error (MAPE) values are reported on the upper right corner of the error maps. Note the enhanced accuracy which is obtained with the optimized flip-angle train especially for $T_2$ reconstructions.

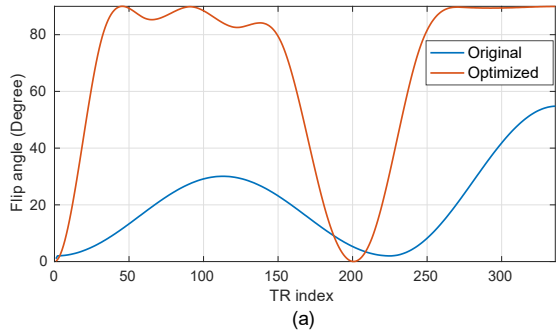

(a)

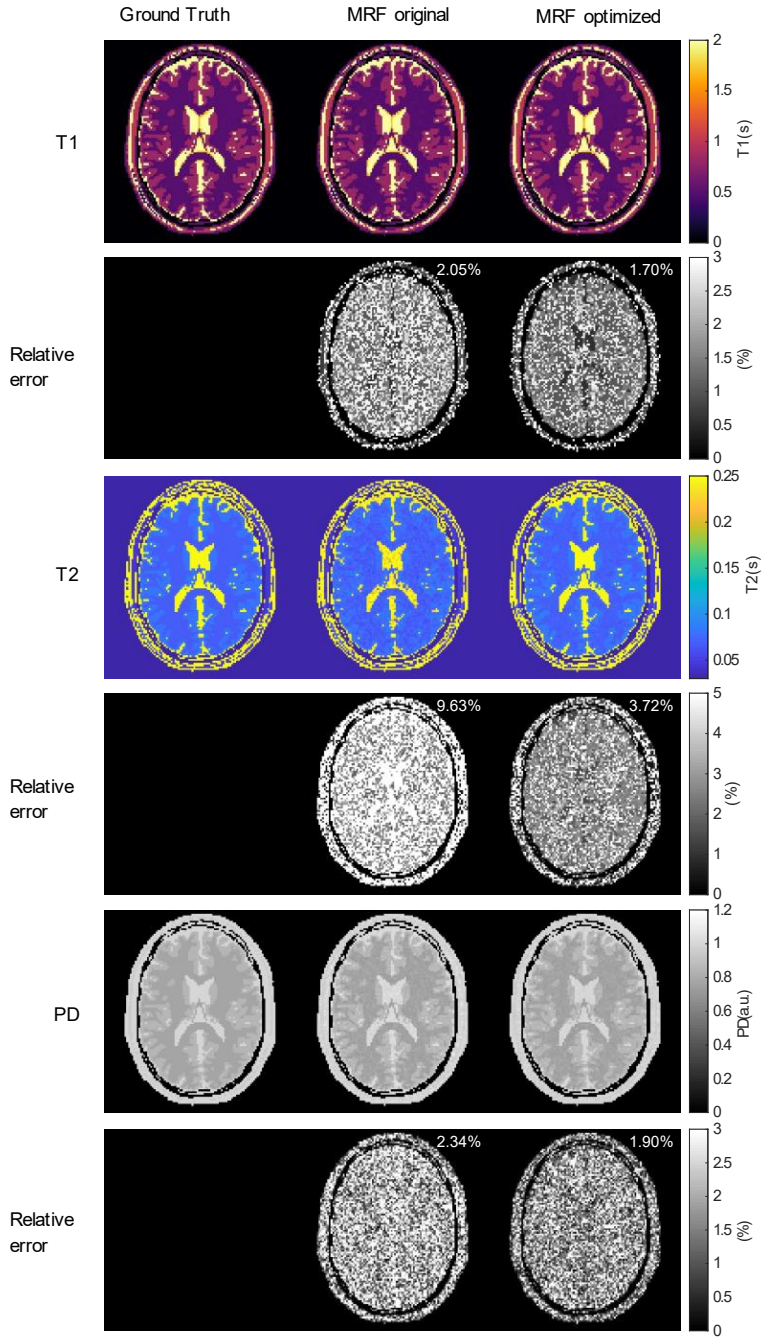

(b)

## Appendix -- Supplementary material

### A) Transforming linear physical operator from Bloch equation into the EPG model

Below is a proof showing that any linear physical operator (e.g., RF excitation computation) used in Bloch simulator can also be used in the EPG model.

Assume discretized time points $t_0, t_1, t_2, \ldots,$ and rewrite the Fourier transform relationship (1) for time $t_{n+1}$ in a matrix form:

$$\bar{\mathbf{F}}(\mathbf{k}, t_{n+1}) = \int_{\mathbf{V}} \widetilde{\mathbf{M}}(\mathbf{r}, t_{n+1})\, e^{-i\mathbf{k}\mathbf{r}} \mathbf{dr},$$

(A1)

where $\widetilde{\mathbf{M}}(\mathbf{r}, t_{n+1}) = [M_+(\mathbf{r}, t_{n+1}), M_-(\mathbf{r}, t_{n+1}), M_z(\mathbf{r}, t_{n+1})]^T$ is the magnetization signal in a transformed coordinate system.

Note that there is a linear transform relationship between the magnetization $\widetilde{\mathbf{M}}(\mathbf{r}, t_{n+1})$ in a transformed coordinate system and the magnetization $\mathbf{M}(\mathbf{r}, t_{n+1}) = \left[M_x(\mathbf{r}, t_{n+1}), M_y(\mathbf{r}, t_{n+1}), M_z(\mathbf{r}, t_{n+1})\right]^T$ in the Cartesian coordinate system,

$$\widetilde{\mathbf{M}}(\mathbf{r}, t_{n+1}) = \begin{bmatrix} 1 & i & 0 \\ 1 & -i & 0 \\ 0 & 0 & 1 \end{bmatrix} \mathbf{M}(\mathbf{r}, t_{n+1}) = \mathbf{S}\mathbf{M}(\mathbf{r}, t_{n+1}).$$

(A2)

If there exists a linear function $\mathbf{R}[\mathbf{X}] = \mathbf{A}\mathbf{X} + \mathbf{b}$ describing the relationship between the magnetization signal $\mathbf{M}(\mathbf{r}, t_{n+1})$ and $\mathbf{M}(\mathbf{r}, t_n)$ as below,

$$\mathbf{M}(\mathbf{r}, t_{n+1}) = \mathbf{R}[\mathbf{M}(\mathbf{r}, t_n)],$$

(A3)

by plugging (A2) and (A3) into equation (A1), the relationship between $\mathbf{F}(\mathbf{k}, t_{n+1})$ and $\mathbf{F}(\mathbf{k}, t_n)$ is given as below,

$$\bar{\mathbf{F}}(\mathbf{k}, t_{n+1}) = \int_V \mathbf{S}\mathbf{M}(\mathbf{r}, t_{n+1}) \, e^{-i\mathbf{k}\mathbf{r}} d\mathbf{r} = \int_V \mathbf{S}\mathbf{R}[\mathbf{M}(\mathbf{r}, t_n)] \, e^{-i\mathbf{k}\mathbf{r}} d\mathbf{r} = \mathbf{S} \int_V \mathbf{R}[\mathbf{S}^{-1}\widetilde{\mathbf{M}}(\mathbf{r}, t_n)] \, e^{-i\mathbf{k}\mathbf{r}} d\mathbf{r}$$

$$= \mathbf{S} \int_V \left(\mathbf{A}\mathbf{S}^{-1}\widetilde{\mathbf{M}}(\mathbf{r}, t_n) + \mathbf{b}\right) e^{-i\mathbf{k}\mathbf{r}} d\mathbf{r} = \mathbf{S} \int_V \mathbf{A}\mathbf{S}^{-1}\widetilde{\mathbf{M}}(\mathbf{r}, t_n) \, e^{-i\mathbf{k}\mathbf{r}} d\mathbf{r} + \mathbf{S} \int_V \mathbf{b} \, e^{-i\mathbf{k}\mathbf{r}} d\mathbf{r}$$

$$= \begin{cases} \mathbf{S}\mathbf{A}\mathbf{S}^{-1}\bar{\mathbf{F}}(\mathbf{k}, t_n) + \mathbf{S}\mathbf{b}, & \text{if } \mathbf{k} = 0 \text{ ,} \\ \mathbf{S}\mathbf{A}\mathbf{S}^{-1}\bar{\mathbf{F}}(\mathbf{k}, t_n), & \text{otherwise.} \end{cases}$$

$$\text{(A4)}$$

Therefore it is proved that any linear physical operator derived from Bloch equation and applied to magnetization signal $\mathbf{M}(\mathbf{r}, t_n)$ could be similarly applied to the configuration state $\bar{\mathbf{F}}(\mathbf{k}, t_n)$ in the EPG model.

**B) Generation of five different types of flip-angle trains**

$N_{TR}$ is the sequence length;

$mod(A, B)$ computes the remander after division of $A$ by $B$;

$\lfloor A \rfloor$ computes the nearest integer no larger than $A$.

(1) *Spline5:*

$m = 5$, randomly sample $\theta_i$ from $[0,120]$ degree for $i \in [1, m]$.

Compute a cubic spline interpolated curve with data points $(0, \ 0), \left( \left\lfloor \frac{i}{n} N_{TR} \right\rfloor, \alpha_i \right) \dots$ or $i \in [1, m]$. Set first-order boundary condition at both ends to be 0.

(2) *Spline11:*

Similar as (1), but with $m = 11$.

(3) *SinSquared5:*

$m = 5$, randomly sample $\theta_i$ from $[0,120]$ degree for $i \in [1, m]$. Flip-angle train is computed by:

$$\alpha(n) = \theta_i \sin^2 \left( \pi \frac{i}{N_{TR}/m} \right), i = mod(n, \lfloor N_{TR}/m \rfloor), for \ n = 1 \dots N_{TR}.$$

The generated flip-angle train has 5 sine square lobes with different amplitudes.

(4) *SplineNoise11*: *Spline11* from (2) plus Gaussian random noise with mean=0 and variance=10.

(5) *PieceConstant5*:

$m = 5$, $N_{min} = 20$.

Random sample integers $k_i$ in an ascending order from $[N_{min}, N_{TR} - N_{min}]$ for $i = 1, 2, \dots, m - 1$. $k_0 = 1, k_m = N_{TR}$.

Inequality $|k_i - k_j| \geq N_{min}$ should hold for $i, j = 0, 1, 2, \dots, m$.

Randomly sample $\theta_i$ from $[0,120]$ degree for $i \in [1, m]$. Flip-angle train is computed by:

$$\alpha(n) = \theta_i, for \ n \in (k_i, k_{i+1}], i = 0, 1, \ldots m-1.$$

The generated flip-angle train has 5 constant-value segments with different amplitudes, with each segment a minimum length of at least 20.